%% file: ms.tex
\documentclass[acmtog,table]{acmart}
\pdfoutput=1

\definecolor{redcolor}{rgb}{0.8,0,0}

\usepackage[ruled,vlined]{algorithm2e}
\usepackage{wrapfig}
\usepackage{enumitem}
\usepackage{nicefrac}
\usepackage{amsmath,amsfonts,amssymb}
\usepackage{breakcites}
\newcommand{\ra}[1]{\renewcommand{\arraystretch}{#1}}
\usepackage{tabularx}
\usepackage{booktabs} 
\usepackage{isomath}
\definecolor{lightbluishgrey}{rgb}{0.78,0.86,0.93}

\usepackage[ruled]{algorithm2e} 

\SetAlFnt{\small}
\SetAlCapFnt{\small}
\SetAlCapNameFnt{\small}
\SetAlCapHSkip{0pt}

\begin{document}

\title{Convolutional Humanoid Animation via Deformation}

\author{John Kanji}
\affiliation{%
  \institution{University of Toronto}
  \country{Canada}}
\email{jkanji@dgp.toronto.edu}

\author{David I. W. Levin}
\affiliation{%
  \institution{University of Toronto}
  \country{Canada}}
\email{diwlevin@cs.toronto.edu}

\renewcommand{\shortauthors}{Kanji and Levin}

\begin{abstract}
\input{abstract}
\end{abstract}



\begin{teaserfigure}
  \centering
  \includegraphics{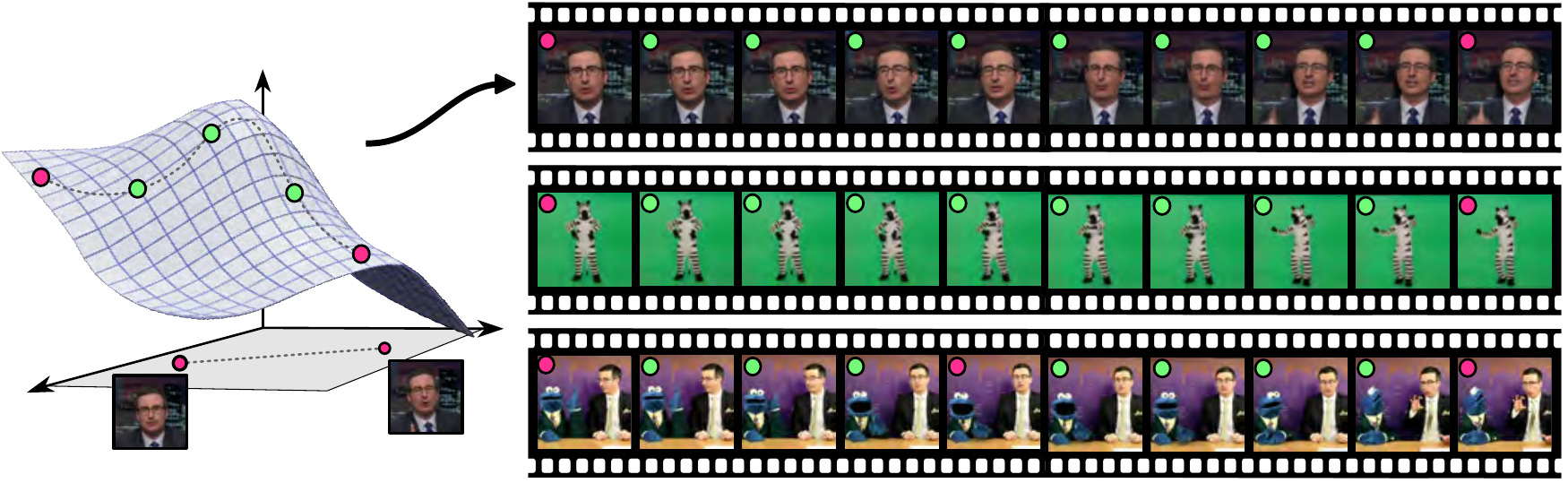}
  \caption{Our Convolutional algorithm for Humanoid Animation via Deformation (CHAD) parameterizes object pose via a learned configuration manifold. CHAD generates new animations (green dots) by following interpolating curves between keyframes (red dots) on this manifold. CHAD uses no prior on motion or subject type, enabling the synthesis of face motion, whole body motion or even multiple character motion.}
\end{teaserfigure}

\maketitle

\input{intro}

\input{lit}
\input{problem}
\input{method}

\input{results}
\input{conclusions}

\bibliographystyle{ACM-Reference-Format}
\bibliography{chad}
\end{document}

%% file: abstract.tex
In this paper we present a new deep learning-driven approach to image-based synthesis of animations involving humanoid characters. Unlike previous deep approaches to image-based animation our method makes no assumptions on the type of motion to be animated nor does it require dense temporal input to produce motion. Instead we generate new animations by interpolating between user chosen keyframes, arranged sparsely in time. Utilizing a novel configuration manifold learning approach we interpolate suitable motions between these keyframes. In contrast to previous methods, ours requires less data (animations can be generated from a single youtube video) and is broadly applicable to a wide range of motions including facial motion, whole body motion and even scenes with multiple characters. These improvements serve to significantly reduce the difficulty in producing image-based animations of humanoid characters, allowing even broader audiences to express their creativity.

%% file: intro.tex
\section{Introduction} \label{intro}
Character animation is hard. In computer graphics, animating a character begins with choosing a suitable motion parameterization called a rig. Rigs come in many forms, from skeletal rigs used to represent human motion to blendshapes, with much inbetween. Even when availing oneself of state-of-the-art approaches to help automate rig construction, the process can be tedious and requires copious amounts of skill and precision. A poorly built rig could be hard to control, exclude important character motions or both. For a blockbuster movie, rigs are works of art, the result of the collaboration of many expert modelers and rigging artists. 

With rig in hand, the real work begins: synthesizing character motion by crafting a time varying trajectory through the rig-space. This process can be artist-guided, driven by motion capture data or even video. However, manually constructing appealing character motions requires patience and an artistic eye for the subtleties of human motion, motion capture requires expensive additional hardware and software and methods for video often rely on strong priors, meaning that no single method is broadly applicable to all types of character animation. 

One approach to ease the burden of character animation is to use image-based approaches. This is quickly becoming a defacto standard approach for facial animation~ and has been explored for character animation as well~. The advantage of these approaches is that, by leveraging machine learning techniques, compelling subspaces for pose can be created and then driven using video, permitting easy synthesis of animations. 

Unfortunately, these approaches typically require strong priors on the poses they can create. Ironically these methods rely on the blendshapes and rigs that are so burdensome in traditional computer animation approaches. A consequence of this is that these algorithms do not apply to general humanoid character motion synthesis -- instead they individually specialize towards face, hand  or body animation~ and will fail for novel, unexpected motions. For instance, systems that rely on face models and facial landmarks can only reproduce poses of the human face. 

Our goal is to produce a general, image-based algorithm for character motion synthesis. In contrast to previous methods our approach for generating convolutional humanoid animation via deformation (CHAD) does so without requiring any explicit prior on the motion type. Instead, we learn an ``implicit rig'' by constructing a configuration space for a particular animation from short videos (we use mostly YouTube videos). 

CHAD's goal is to provide a general purpose tool that will allow inexperienced users to craft new image-based animations via keyframing. CHAD requires comparatively little data and paths in the CHAD configuration space encode natural human movements (replete with hand wringing, facial ticks and blinks) meaning that, with relatively little input (a few keyframes), a novice can synthesize a compelling animation. While CHAD does not match the highest quality animations produced by professionals, its ease-of-use, expressiveness and ability to generate a wide range of varying motions make for a significant step towards the democratization of quality image-based animation.

%% file: lit.tex
\section{Related Work} \label{lit}
The goal of synthesizing humanoid motion drives a large portion of the computer graphics research community. An exhaustive characterization of all related work is beyond the scope of this paper. Below, we attempt to highlight important developments and position our work, CHAD, appropriately relative to this ever growing corpus. 

Of all the types of humanoid animation to be studied, that of the face has seen, perhaps, the most attention. Everything from highly detailed facial capture~\cite{Beeler2010,Beeler2011} to sensorimotor modeling~\cite{Lee:2006:HUB:1141911.1142013} has been employed to generate convincing facial animations. 

Facial motion capture lies at the heart of a large number of face animation algorithms and has become increasingly popular as both a research and industrial tool. Williams~\shortcite{Williams:1990:PFA:97880.97906} introduced the notion of marker-based 3D face capture, while the seminal work of Bradley et al.~\shortcite{Bradley2010} debuted a markerless approach which relies on multiview stereo to fit geometry from images, combined with optical flow to track  deformation and texture details across frames. Beeler et al.~\shortcite{Beeler2011} improve this method by introducing anchor frames to track facial motion while avoiding integrated error. An increasingly large number of facial animation papers rely on face capture to provide input data for data-driven approaches.

A classical approach to data-driven facial animation is the so-called 3D Morphable Model (3DMM)~\cite{Blanz1999}. This method models textured 3D faces from data using principal component analysis (PCA) to project captured data into a low-dimensional parameter space. The model can then be controlled by fitting the parameters to a photograph. Creating an animation thus requires a dense sequence of control frames. By the nature of the PCA projection, fine-scale pose details are lost in the reconstructed model. However, this approach has formed the basis of a several followup works such as Kim et al.~\shortcite{kim2018deep} or Olszewski et al.~\shortcite{Olszewski2017}, who use 3DMM to model the poses of a source video, and then transfer these motions to a target portrait, relying on the source video to provide fine details. These methods can be considered image-based approaches that rely on a strong facial prior. They produce compelling results but are limited to faces only and can have difficulty with features such as long hair, which the 3DMM prior does not model.

Blendshapes provide an alternative reduced space representation for facial motion synthesis. Many approaches use dense temporal input such as monocular images~\cite{Cao2014,Cao2016} or even strain gauge data~\cite{li2015facial} to drive blendshape models and produce 3D facial animations. While blendshapes can represent complicated facial expressions more compactly then PCA-based 3DMMs, they still can exclude detailed motions and are often augmented at runtime to make up for this~\cite{Cao2015}. Accurate interpolation between tracked expressions can also be difficult. Meng et al.~\shortcite{Meng2018} tackle this by learning an embedding of facial expression from images, which forms the input to a recurrent model giving 3D deformations between face poses. 

The related problem of facial motion transfer has also seen much interest. Xu et al~\shortcite{Xu2014} seek to transfer the 3D motions from a source model to a target. Large-scale motions are transferred using a blendshape model and deformation transfer~\cite{Sumner2004DeformationTF}, with fine details transferred using the coating transfer method~\cite{SorkineHornung2004LaplacianSE}. This method requires temporally dense 3D mesh input which can be cumbersome to acquire, process and store. Garrido et al.~\shortcite{Garrido2015VDubMF} apply motion transfer to dialog dubbing, transferring the mouth movements of a dubber onto an actor's performance. They also employ a blendshape model derived from monocular facial capture as a prior on facial expression. Finally, Vlassic et al.~\shortcite{Vlasic2006} employ a multilinear model to factorize the variance of a set of 3D poses into identity, expression, and viseme. They can then transfer appearance by traversing the identity axis of the reduced space. 

Finally, 3D facial motion synthesis cannot be discussed without at least some discussion on animation of speech. Edwards et al.~\shortcite{Edwards2016JALIAA} introduce an artist friendly Jaw-Lip space to apply and control lip-synchronized animation synthesis, based on audio and textual input. Suwajanakorn et al.~\shortcite{suwajanakorn2017synthesizing} take a deep learning approach, using a recurrent neural network to synthesize an appropriate mouth texture from audio input, allowing the generation of speech. This method uses facial landmarks as a prior to guide the lip-synch process.

While the previously mentioned works treat facial motion in the 3D domain, others have used image-based approaches, analyzing and manipulating motion in 2D pixel space. Garrido et al.~\shortcite{Garrido2014AutomaticFR} use identity-preserving image warps to perform facial reenactment and use a face-matching metric to chose frames to be transferred from a source clip to a target performance. Other approaches blend between different recorded takes of the same performance~\cite{Malleson2015}, generate cinematographs~\cite{aberman2018neural} or animate portraits~\cite{Averbuch-Elor2017,geng2018warp} using deformation fields. 

Beyond faces, full body human motion has also been explored extensively and again can be split into the broad categories: 3D and image-based approaches. Skeletal rigs are a popular choice for parameterizing human motion and there has been extensive work exploring both performance capture\cite{Xu2018MonoPerfCapHP}, motion synthesis~\cite{Holden2015LearningMM,Holden2016ADL,Holden2017PhasefunctionedNN} and control~\cite{Peng2017DeepLocoDL,Peng2018DeepMimicED,Yu2018LearningSA} of such representations.
Image-based method have also been explored for whole body animation. Chan et al.~\shortcite{Chan2018EverybodyDN} perform motion transfer of dance motion, using a learned 2D pose estimator, while other works\citet{Davis:2018:VRB:3197517.3201371} re-time video to synchronize with the beat of a user-supplied song. Other methods synthesize images of novel poses of humans from video input~\cite{Balakrishnan2018SynthesizingIO} or learn to predict future frames in a video sequence~\cite{Xue2016VisualDP,aberman2018neural}.

 Almost all previous approaches for image-based animation rely on strong priors to generate results (such as 3DMMs or blendshapes for facial animation, or a known skeleton for full humanoid motion). General, image-based approaches for approximating shape variation~\cite{Cootes1995} and tracking motion~\cite{Lucas1981AnII,Tao2012SimpleFlowAN,Weinzaepfel2013DeepFlowLD,Grover2015liteFlowLA} have been developed but, when applied to the problem of animation, they either sacrifice detailed motion (PCA-based approaches) or break down for long running trajectories (optical flow based approaches). In contrast to all these methods, CHAD  eschews any prior information regarding the structure of the character to be animated. Rather CHAD learns an ``implicit rig'' for a face, character or multiple characters entirely from a small amount of data (we rely on single YouTube videos for most of the examples in this paper). Rather than relying on dense temporal input, CHAD can interpolate natural looking motion between sparsely placed keyframes, making animator control intuitive. In the next sections we detail the asymmetric neural net that lies at the heart of CHAD. 



%% file: problem.tex
\section{Problem Overview} \label{problem}
\begin{figure*}[htp]
	\includegraphics[width=\textwidth]{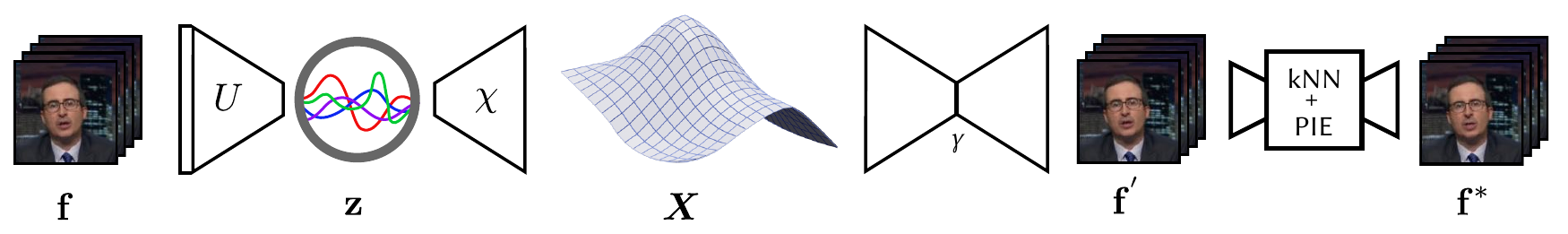}	
	\caption{CHAD-Net. Our asymmetric network setup for learning cool animations. Our method takes, as input, a set of video frames ($\mathbf{f}$) and passes them through an assymetric autoencoder where the encoder $U$ is a single layer composed of the PCA-basis and $\mathit{X}$ is a deep, convolutional decoder. Using this architecture we learn a configuration manifold ($\boldsymbol{X}$) that encodes the video motion. Concurrently we train a GAN ($\gamma$) that produces low resolution images  which we further improve using detail transfer from the initial set of input frames. }
	\label{fig:chadNET}
\end{figure*}

CHAD (\autoref{fig:chadNET}) aims to place as few restrictions on input as possible, preferring to ingest unlabeled, unstructured videos;
indeed, many of the examples shown here have been scraped from YouTube,
and all are unprocessed aside from uniform cropping.
We chose keyframe-based animation as our control method because it provides an intuitive interface for the user to
specify target poses while also being familiar to experienced animators. It also provides flexibility for the user
to choose the granularity or sparsity of their keyframes. The output of CHAD is a video sequence that interpolates between user specified keyframes with motion that mimics that of the input video sequence. 

%% file: method.tex
\section{Method} \label{method}
In this section we detail both the experiments and insights that led to the development of CHAD, as well as the workings of CHAD itself. 
\input{deformations}
\input{manifold}
\input{synthesis}




%% file: deformations.tex
\subsection{Deformation Field Learning} \label{sec:deformations}

Our initial attempts at learning a model for humanoid animation were inspired by previous work on video frame prediction. These algorithms often learn deformation fields between adjacent frames in a video sequence. A deformation field, $\boldsymbol{u}^i\in\mathbb{R}^{2n\times 2n}$, is a two-dimensional vector field over an $n\times n$ image. This field encodes the deformation of the $i^{th}$ image in a video sequence ($\boldsymbol{f}^{i}$) into the $({i+1})^{th}$ frame ($\boldsymbol{f}^{i+1}$). In practice, one can reconstruct $\boldsymbol{f}^{i+1}$ via a deformation operation $\mathcal{D}$ such that $\boldsymbol{f}^{i+1} = \mathcal{D}\left(\boldsymbol{f}^{i}, \boldsymbol{u}^{i} \right)$. Typically $\mathcal{D}$ is a bilinear image warp. 

Creating a user controlled animation requires, at the very least, the ability to specify a starting state for the animation and then to be able to evolve that state over time. We can accomplish this by beginning with a suitable initial frame $\boldsymbol{f}^0$ and progressively warping it with displacement fields that characterize the motion desired for the animation. An obvious approach to parameterizing this space of displacement fields is to learn a reduced mapping from exemplar data, via Deep Learning. 

\begin{figure}[H]
	\includegraphics[width=\columnwidth]{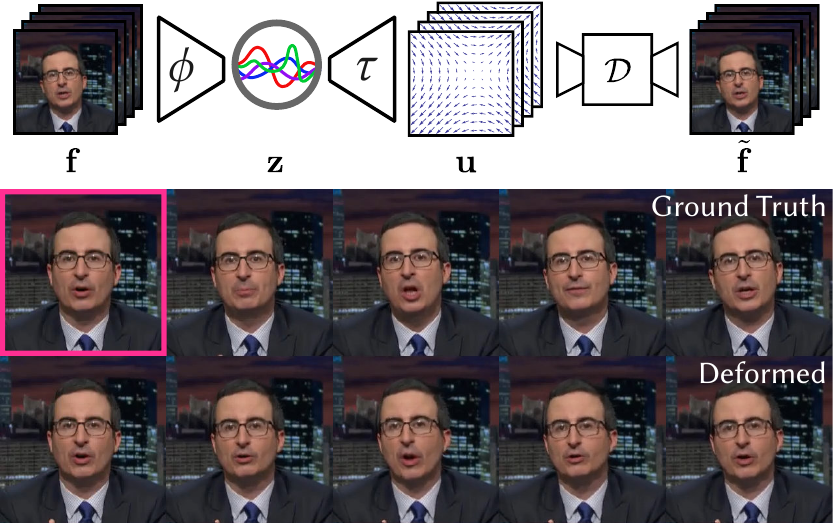}	
	\caption{Deformation field learning setup. The ground truth is reconstructed by deforming the reference frame.}
	\label{fig:defNN}
\end{figure}

The input to our deformation learning algorithm is an input video sequence, $\mathit{F}$ (e.g. of a human face speaking).  We construct a convolutional auto-encoder by composing an encoder: $\phi : \mathit{F} \to \mathit{Z}$ and a decoder: $\tau : \mathit{Z} \to \mathit{U}$. Here $\mathit{U}$ is the space of all displacement fields and $\mathit{Z}$ is a reduced space with $|\mathit{Z}| << |\mathit{U}|$. This encoder-decoder pair is parameterized by a pair of convolutional neural networks (CNNs)~(\autoref{fig:defNN}).

Our loss function is crafted such that $\tau$ learns to produce deformation fields that warp frames $\boldsymbol{f}^i$ to frames $\boldsymbol{f}^{i+1}$. The encoder-decoder pair can be trained to do this by minimizing the following loss,
\begin{equation}
	\mathcal{L}_\text{def} = \left|\boldsymbol{f}^{i+1} - \mathcal{D}\left(\boldsymbol{f}^{i}, \tau(\phi(\boldsymbol{f}^i, \boldsymbol{f}^{i+1})) \right)\right|_1,
	\label{eq:deformationCost}
\end{equation} 
for each $\boldsymbol{f}^i, \boldsymbol{f}^{i+1}\in \mathit{F}$. 
Full details of the training procedure are given in \autoref{training}. Once trained, we can synthesize the $k^{th}$ frame of a new animation by evaluating a sequence of $k$ deformations using our learned displacement fields. However, this approach fails in practice due to the accumulation of error caused by succesive warping.
	
\begin{figure}[H]
	\includegraphics[width=\columnwidth]{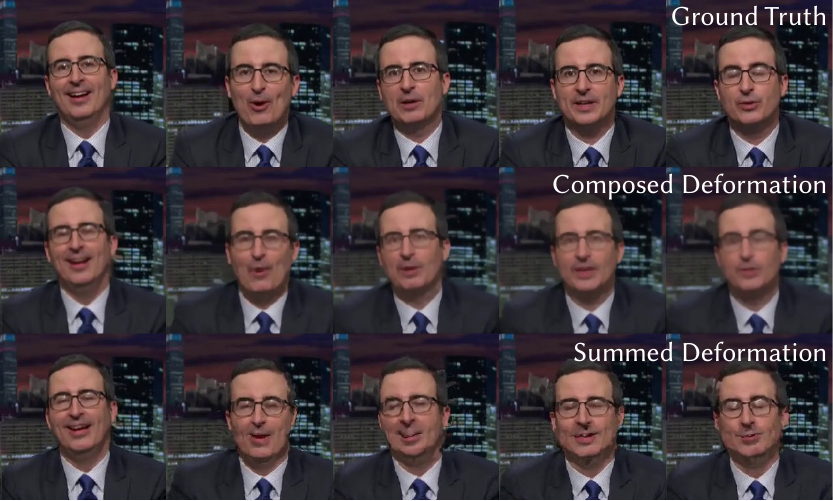}	
	\caption{Accumulated error incurred over 5 seconds by our two deformation methods.}
	\label{fig:defNNResults}
\end{figure}

To combat error accumulation, we train using batches of sequential frames, attempting to reconstruct the sequence from the first frame in the batch. To reconstruct the $i^{th}$ frame in the sequence we employ two methods, summed deformations and composed deformations.
For summed deformation we compute $\boldsymbol{f}^i$ by
\begin{equation}
	\boldsymbol{f}^{i+1} = \mathcal{D}\left(\boldsymbol{f}^0, \sum_{j=0}^i \tau \left(\phi\left(\boldsymbol{f}^j\right)\right)\right).
	\label{eq:summedDef}
\end{equation}
Composed deformations are given by
\begin{equation}
	\boldsymbol{f}^{i+1} = \mathcal{D}_{i}^L\circ\cdots\circ\mathcal{D}_0^L\left(\boldsymbol{f}^{0}\right),
	\label{eq:deformationCost2}
\end{equation} where we take $\mathcal{D}_i^L$ to be our learned deformation function, given by $\mathcal{D}\left(\boldsymbol{f}^i, \tau(\phi(\boldsymbol{f}^{i}, \boldsymbol{f}^{i+1})\right)$ while $\circ$ is the standard function composition operator. These two methods accumulate error differently (as can be seen in \autoref{fig:defNNResults}).
To balance out these errors, we use a deformation loss that is the sum of the $L_1$ distances computed by reconstructing the batch using each method.

In practice we observed two problems with this model. The first was that, while it could readily fit to our training data and produce a plausible reconstruction, synthesizing animations via repeated applications of $\mathcal{D}^L$ produced poor results. The second was that when learning the deformation between frames, the input space of the encoder is large; it's $F\times F$. The autoencoder needs to see a lot of data to fully characterize this mapping. Second, while the deformation learning approach allows us to grow an animation out from an initial frame, it does not readily allow us to interpolate between two separate key frames. This makes animations produced in this manner extremely difficult to control for a user. These two observations led us to incorporate more structure into CHAD via the use of configuration manifolds, which we detail below.

%% file: manifold.tex
\subsection{The Configuration Manifold} \label{manifold}

Deformation fields are a very general motion representation. Our goal is not to represent any motion in a video but to generate new videos that contain humanoid motion. To address the issues discussed above we instead turn to a representation common in the field of mechanics~\cite{lanczos1986variational}. In mechanics, the set of all poses of an object are modeled by a low-dimensional manifold (the \emph{configuration manifold}) embedded in high-dimensional space.  

Each point, $\boldsymbol{x}\left(\boldsymbol{z}\right)$,  on a \emph{configuration manifold}, $\boldsymbol{X}$, represents a unique pose of an object. Here  $\boldsymbol{x}$ is not a point in 3D space but an n-dimensional point that describes the deformed state of the object (e.g. for a triangle mesh  $\boldsymbol{x}$ stores all vertex positions of the mesh), while $\boldsymbol{z}\in\mathit{Z}$ is a low-dimensional coordinate (the rotation matrix and translation vector of a rigid body, for instance). In our case, we have no explicit knowledge of the object's form. However, we can attempt to learn a proxy to $\mathit{X}$ from input video. 

Any smooth, continuous motion of an object can be described by a corresponding smooth continuous path on $\mathit{X}$. Given a time varying motion $\boldsymbol{x}\left(t\right)$, we note that the object's instantaneous velocity is given by $\frac{d\boldsymbol{x}}{dt}=\frac{\partial \boldsymbol{x}}{\partial \boldsymbol{z}}\frac{\partial \boldsymbol{z}}{\partial \boldsymbol{t}}$ and that, over a sufficiently small time interval $\Delta t$ we can represent the displacement of every point in an object as 
	
	\begin{equation}
		\boldsymbol{u} = \frac{d\boldsymbol{x}}{dt}\Delta t = \frac{\partial \boldsymbol{x}}{\partial \boldsymbol{z}}\frac{\partial \boldsymbol{z}}{\partial \boldsymbol{t}}\Delta t, 
		\label{eq:manifoldDisp}
	\end{equation} which is the standard relationship between an objects total velocity and its velocity in the reduced space $\boldsymbol{Z}$.

In CHAD, we take $\boldsymbol{x}$ to be an $n \times n$ pixel image and we take $\Delta t$ to be the frame time of our input video (typically $\frac{1}{30}$ of a second). We can now use \autoref{eq:manifoldDisp} to establish the following relationship between $\mathit{F}$ and $\mathit{X}$:
	\begin{equation}
		\begin{split}
			\boldsymbol{u}^{i+\frac{1}{2}}  =  \boldsymbol{x}^{i+1} - \boldsymbol{x}^{i} \\
			\boldsymbol{f}^{i+1} = \mathcal{D}\left(\boldsymbol{f}^{i}, \boldsymbol{u}^{i+\frac{1}{2}}\right),
			\label{eq:manifoldFD}
		\end{split}
	\end{equation} via finite differences. In order to simplify the equation, we take the distance over which the finite difference is calculated to also be $\Delta t$ (which is the smallest unit of time we can observe from our input videos). The $\frac{1}{2}$ frame increment denotes that $\boldsymbol{u}^{i+\frac{1}{2}}$ is being estimated at the midpoint of the line between $\boldsymbol{x}^{i}$ and $ \boldsymbol{x}^{i+1}$. 

\begin{figure}[h]
	\includegraphics[width=\columnwidth]{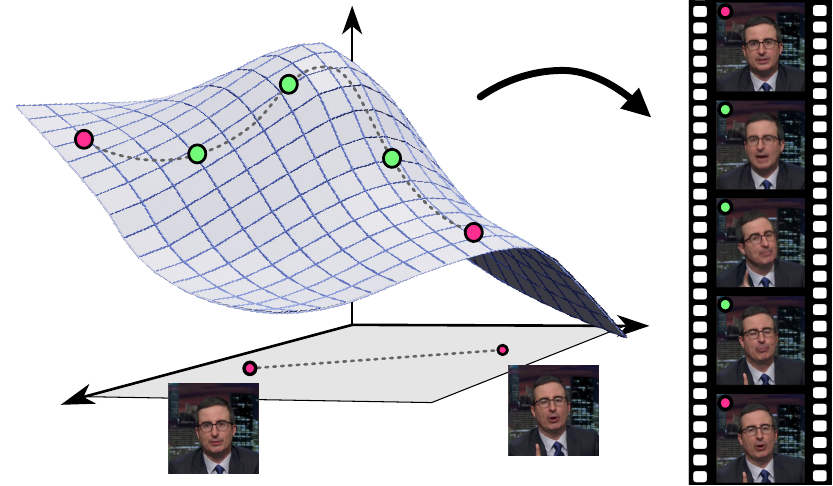}	
	\caption{A hypothetical configuration manifold for image-based animation. Our algorithm maps from a low dimensional space to the high-dimensional pose manifold. Animation sequences (green dots) are curves on the surface of this manifold that interpolate between user chosen keyframe (red dots)}
	\label{fig:manifoldJO}
\end{figure}

\autoref{fig:manifoldJO} shows a hypothetical configuration manifold for image-based animation. The configuration manifold representation addresses the two issues with deformation field learning. First, the manifold is more constrained than the mapping learned in \autoref{sec:deformations}. Displacements emanating from identical frames are compactly encoded in the tangent space of the manifold whereas deformation learning maps each of these to a unique point in the reduced space~(\autoref{fig:deepManifold}). Second, interpolating between two animation frames becomes as easy as following a curve between them in $\mathit{Z}$ and retrieving the image frames via the mapping $\boldsymbol{x}\left(\boldsymbol{z}\right)$.

\subsubsection{Deep Configuration Manifolds} 
\begin{figure}[H]
	\includegraphics[width=\columnwidth]{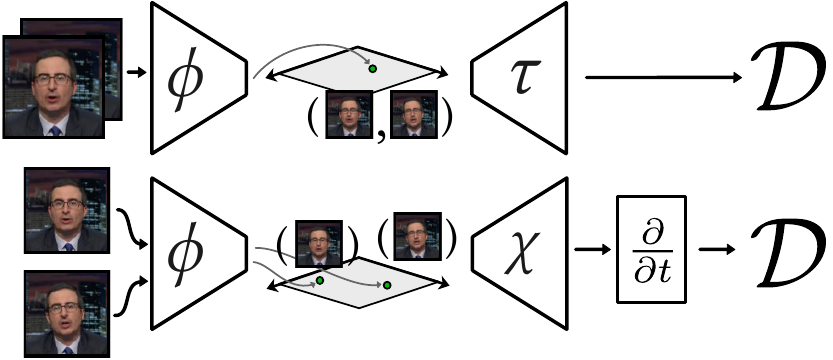}	
	\caption{Deformation learning (top) requires two frames as input, giving a very large space to learn. Using the configuration manifold we can project each frame into a common space (bottom) and compute curves between them.}
	\label{fig:deepManifold}
\end{figure}

We will modify our Deep Learning approach from ~\autoref{sec:deformations} to learn the configuration manifold. This requires replacing $\tau$ with $\chi : \mathit{Z} \to \boldsymbol{X}$ (\autoref{eq:deformationCost}). Composition with $\phi$ gives $\chi \circ \phi : \mathit{F} \to \boldsymbol{X}$, which projects a frame onto the configuration manifold. Enforcing the structure imposed by \autoref{eq:manifoldFD} is done, not by modifying \autoref{eq:deformationCost}, but by modifying the structure of the autoencoder itself~(\autoref{fig:deepManifold}). 

We learn a CNN encoder-decoder pair, $\phi : \mathit{F} \to \mathit{Z}$ and $\chi : \mathit{Z} \to \boldsymbol{X}$,
which projects a frame into a low dimensional pose space $\mathit{Z}$, and then onto the configuration manifold.
We train the network by considering a batch of sequential frames,  in $\mathit{F}$.
By mapping the sequence onto the configuration manifold we obtain a piece-wise linear curve on the manifold. For each segment of this curve between frames $\boldsymbol{f}^i$ and $\boldsymbol{f}^{i+1}$, we compute $\boldsymbol{u}^i$ using \autoref{eq:manifoldFD} where $\boldsymbol{x}^i$ (resp. $\boldsymbol{x}^{i+1}$) is the current estimate for the configuration point of $\boldsymbol{f}^i$ (resp. $\boldsymbol{f}^{i+1}$).

\subsubsection{A Step Too Deep?}
 \label{man-too-deep}
\begin{figure}[H]
	\includegraphics[width=\columnwidth]{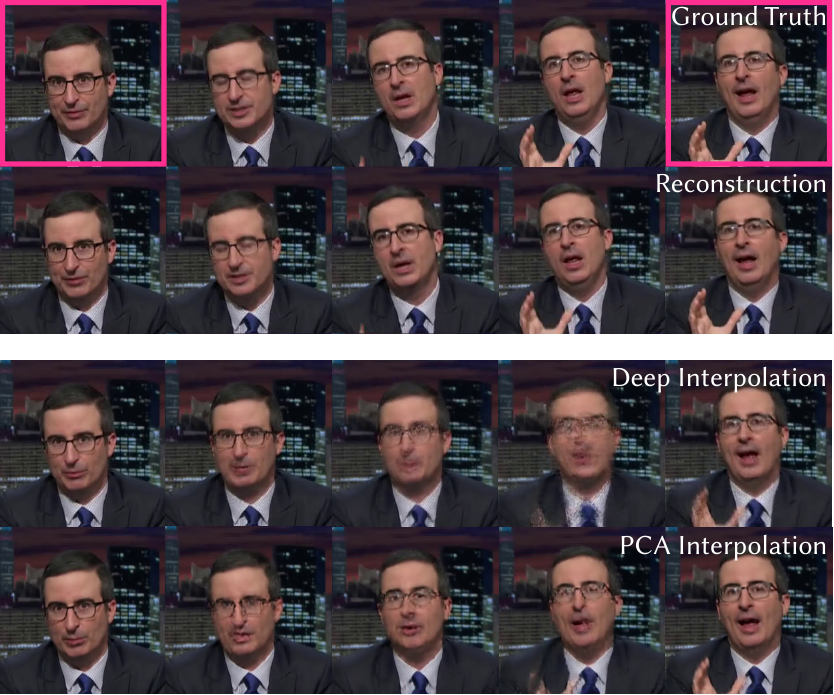}	
	\caption{Top: Ground truth video sequence. Second: Reconstruction using Deep Encoder-Decoder. Third: New sequence synthesized by interpolating between end points using Deep Encoder-Decoder. Bottom: Interpolation using PCA Encoder-Deep Decoder. Notice how the PCA encoder produces a more detailed result.}
	\label{fig:encoderCompare}
\end{figure}

Once we have learned the configuration manifold we can embed any two keyframes into $\boldsymbol{Z}$ and perform interpolation by walking a smooth curve between them. We can hallucinate new frames either using $\mathcal{D}^L$ or more advanced methods~(\autoref{sec:synthesis}). Interpolation performs well for frames that were temporally close in the input video, but the results degrade quickly as the keyframes get further apart, both in terms of visual fidelity, and quality of motion~(\autoref{fig:encoderCompare}).

Let's consider the effect of a small perturbation on a single frame of an animation on its configuration point:
\begin{equation}
	\boldsymbol{x}\left(\boldsymbol{f}+\Delta\boldsymbol{f}\right) \approx \boldsymbol{x}\left(\boldsymbol{f}\right) + \underbrace{\frac{\partial \boldsymbol{x}}{\partial \boldsymbol{z}}}_{\mbox{\autoref{eq:manifoldFD}}}\frac{\partial \boldsymbol{z}}{\partial \boldsymbol{f}}\Delta\boldsymbol{f}.
	\label{eq:imPertubation}
\end{equation} From this simple, local expansion it is easy to see that the inclusion of the deep encoder $\boldsymbol{z}\left(\boldsymbol{f}\right)$ potentially allows an image perturbation to cause a finite, but unbounded perturbation in the corresponding configuration coordinate. This is because, while $\frac{\partial \boldsymbol{x}}{\partial \boldsymbol{z}}$ is regularized by \autoref{eq:manifoldFD} (which we attempt to enforce via deformation loss), $\frac{\partial \boldsymbol{z}}{\partial \boldsymbol{f}}$ has no such regularizer. 

Ideally, the matrix 2-norm of $\frac{\partial \boldsymbol{z}}{\partial \boldsymbol{f}}$ should be bounded with a maximum value of $1$. so that as much of the pose change induced by an image perturbation is included in $\boldsymbol{x}\left(\boldsymbol{z}\right)$ as possible. A simple method for constructing such a space $\mathit{Z}$ is to apply PCA to our input frame data. The resulting linear basis, $\mathbf{Z}$, can be represented as an orthogonal matrix. Using $\mathbf{Z}^T$ as our encoder gives us the property we want. \autoref{fig:encoderCompare} shows the improvement that can be gained by using a PCA encoder and \autoref{fig:chadNET} shows the final, asymmetric autoencoder setup used to generate all subsequent results.

%% file: synthesis.tex
\subsection{Image Synthesis} \label{sec:synthesis}
Until this point we have assumed new images are generated via deformation. In some cases this yields acceptable animations but it suffers from the limitation that  we cannot synthesize any visual phenomena not present in the original keyframe being warped. If the objects in the video change their topology (for example, by opening their mouth), this action cannot be reconstructed accurately by deformation alone.

Fortunately, we can leverage our configuration manifold representation to side step this issue. Because, a point on the manifold encodes pose we can learn an image generator $\gamma : \boldsymbol{X} \to \mathit{F}$, which reconstructs a frame based on its pose without relying on the keyframe. This approach has the double benefit of enforcing empirically our requirement of the manifold that it is possible to uniquely reconstruct the frame from its configuration variable.

We implement $\gamma$ as a generative adversarial network (GAN)~\cite{Goodfellow2014GenerativeAN} which takes as input a point on the configuration manifold, and attempts to invert the mapping $(\phi \circ \chi)$, reconstructing the input frame. The GAN's encoder and decoder networks share the architecture of the $\phi$ and $\chi$ networks~(\autoref{fig:encoder} and \autoref{fig:decoder}). The discriminator network uses the encoder architecture with an output vector size of $1$, and a sigmoid function for the final activation layer.

While our image generator gives very good results when reconstructing frames from the video, synthesizing frames via interpolation can introduce noise where the manifold is less well defined. To compensate we perform additional data-driven denoising. 

\subsubsection{Image Denoising via Detail Transfer} 
\label{sec:synth-denoise}
In our experiments we were unable to train our image generator to produce crisp results in all cases. In order to alleviate this problem we turn to classical methods for reconstructing image sequences by transferring detail from existing video frames onto frames synthesized by our image generator. 

Given a noisy synthesized target frame $\boldsymbol{f}_t$, we must first select a source frame from the video whose pose matches the target as closely as possible. We initially choose a set of candidate frames, $\boldsymbol{f_s} \in \mathit{F}$, by projecting $\boldsymbol{f}_t$ into $\mathit{Z}$ and choosing the frames corresponding to the $k$ nearest neighbors~\cite{annoy} by Euclidean distance in $\mathit{Z}$.

\begin{figure}[h]
	\includegraphics[width=\columnwidth]{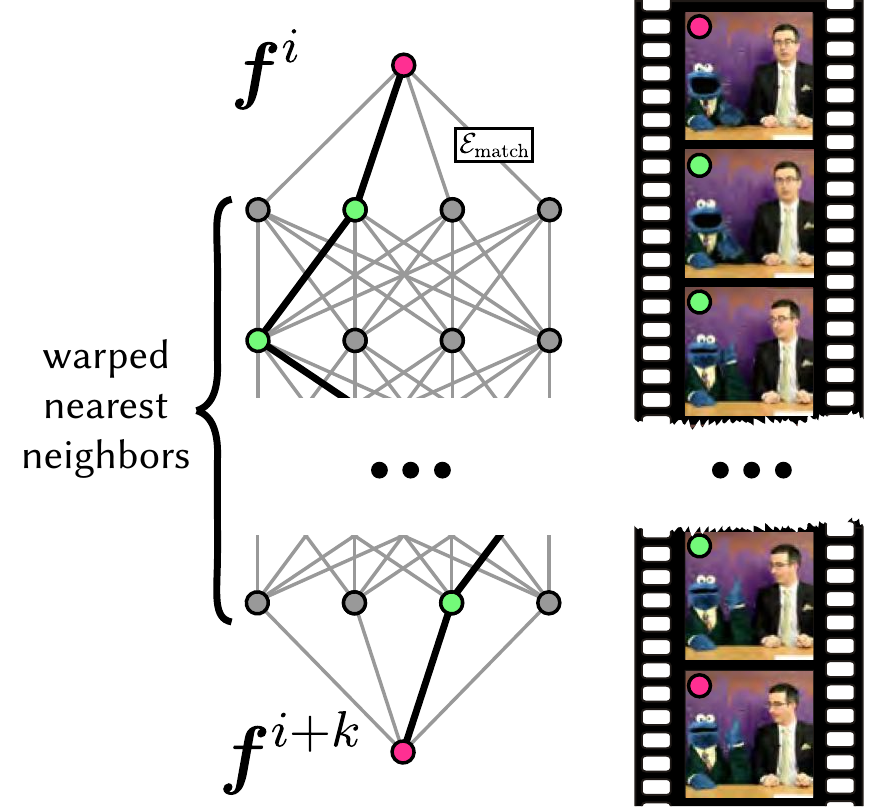}	
	\caption{Left: the graph traversal used for locating frame sequences for detail transfer. Right: an output sequence.}
	\label{fig:denoising}
\end{figure}

\begin{figure}[h]
	\includegraphics[width=\columnwidth]{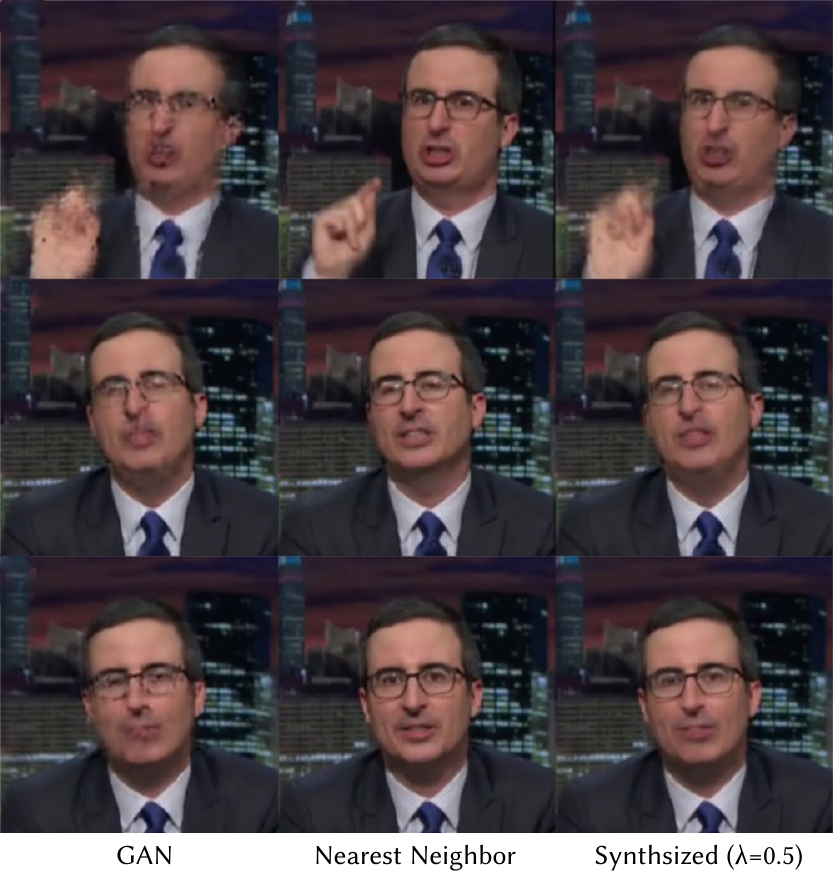}	
	\caption{Examples of our image denoising procedure showing the noisy GAN frame (left), warped nearest 
	neighbor (middle), and final composited frame (right).}
	\label{fig:imageSynthesis}
\end{figure}

We then compute and apply an image warp to each candidate~\cite{Weinzaepfel2013DeepFlowLD} to more closely match the large scale pose in $\boldsymbol{f}_t$, giving $\boldsymbol{\tilde{{f}_s}}$.  We extract an as-smooth-as-possible set of frames using a minimum cost path approach. We construct a directed graph, $E$, in which the source and sink nodes are the keyframes to interpolate between. For each sampled point on the configuration manifold, we add it's $k$ nearest neighbors $E$ (\autoref{fig:denoising}). We set the weights of each edge according to the following cost: 

\begin{figure}[h]
	\includegraphics[width=\columnwidth]{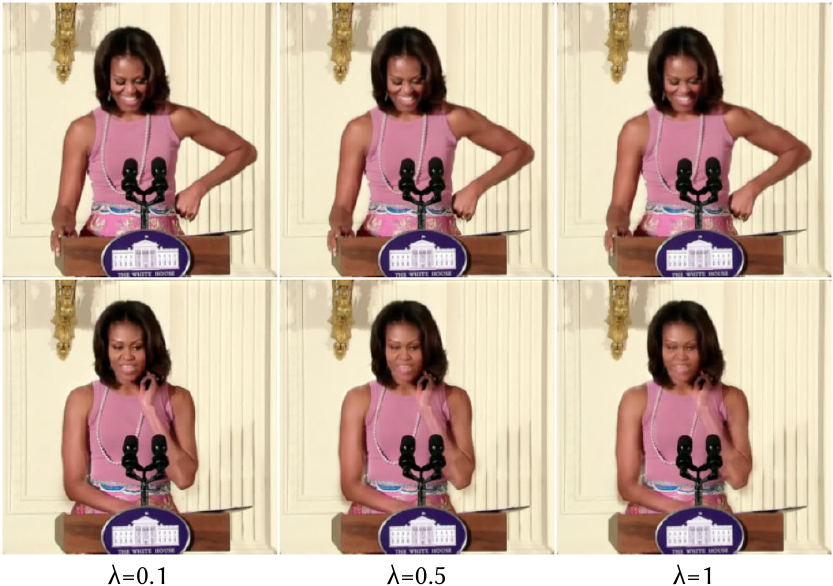}	
	\caption{Increasing the blending parameter lambda, increases the amount of color information taken from  the output GAN frame. }
	\label{fig:lambda}
\end{figure}

\begin{equation}
\begin{aligned}
	\mathcal{E}_{match} =& \alpha\mathcal{E}_\text{data} + \beta\mathcal{E}_\text{smooth} \\
    \mathcal{E}_\text{data} =&\ L_1(\boldsymbol{f}_t^i, \boldsymbol{\tilde{{f}_s}}^i) \\
    \mathcal{E}_\text{smooth} =&\ 
    \begin{cases}
        0,                                                     & \text{if}\ i = 0 \\
        L_1(\boldsymbol{\tilde{{f}_{s}}}^i, \boldsymbol{{f}_{s^*}}^{i-1}),   & \text{otherwise}
    \end{cases}
\end{aligned},
\end{equation} where $\alpha$ and $\beta$  are user-specified parameters, $s^*$ denotes all $k$ nearest neighbors and $i$ is the index (in time) for each frame. The minimum cost path through $E$ gives us a sequence of video frames which we use for detail transfer. We blend details between frames in the standard manner, by solving the screened poisson equation~\cite{Darabi:2012:IMC:2185520.2185578}
\begin{equation}
    (\mathbf{L} + \lambda \mathbf{I})\ \boldsymbol{f}^i = \mathbf{L} \boldsymbol{\tilde{f}}^i_{s} + \lambda \boldsymbol{f^i_t},
\end{equation} where $\mathbf{L}$ is the discrete laplacian for our image, discretized on a regular grid and, $\boldsymbol{f}_t^i$ and $\boldsymbol{\tilde{f}}_{s}^i$ are the target image and warped source image (from the shortest path) in vector form for the $i^{th}$ frame. Intuitively $\lambda$ controls how much of the source frame is included in the final image (\autoref{fig:lambda}). \autoref{fig:imageSynthesis} shows a comparison of images created using the above procedure to frames directly output by $\gamma$.

%% file: results.tex
\section{Evaluation and Results}
\label{sec:results}

\subsection{Training Setup and Data} \label{training}
We implement our neural models using the PyTorch deep-learning framework \citep{paszke2017automatic}.
We use the Adam optimization algorithm \citep{Kingma2014AdamAM} for all models, except for the GAN discriminator,
for which we use stochastic gradient descent in order to stabilize training. 
In all cases the learning rate is set to $1\mathrm{e}{-5}$.
Each minibatch is composed of 32 sequential frames.
We use a progressive training curriculum wherein the training set is periodically expanded.
We begin by seeding the manifold with 100 sequential frames, and train for 50 epochs.
We iteratively expand the training set and train for a further 50 epochs until all frames are included.
The discriminator is trained using binary cross entropy loss using uniformly sampled soft labels of $[0, 0.1]$ for
real images, and $[0.9, 1]$ for synthetic images. Labels are randomly flipped with probability $0.1$ each batch.
Training was performed on various GPUs, noted in Table \ref{tab:data}.

\begin{figure}[h]
	\includegraphics[width=\columnwidth]{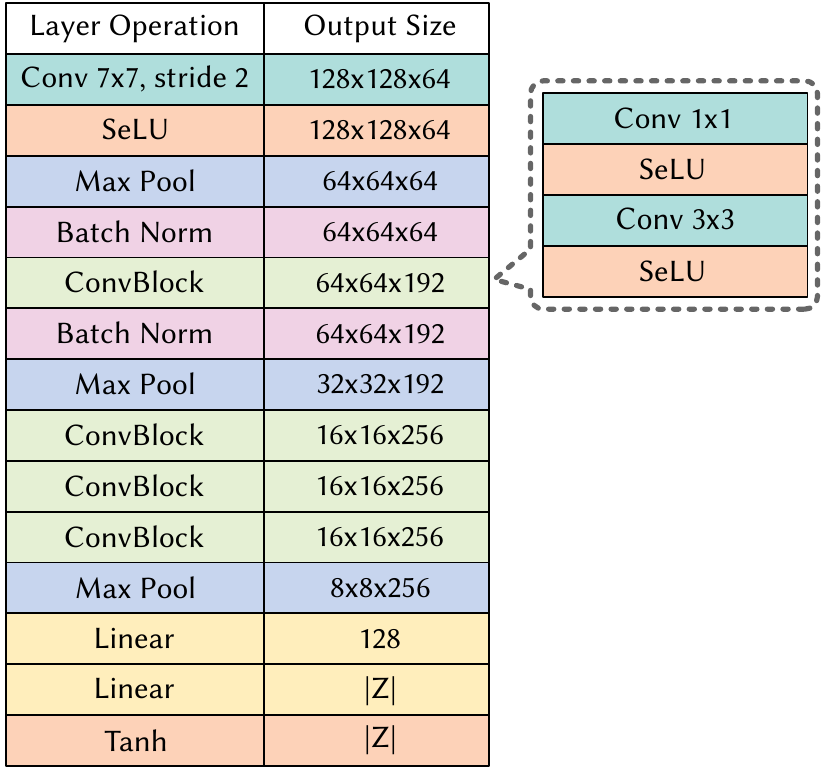}	
	\caption{Encoder architecture. Output sizes are given for input size $256x256$}
	\label{fig:encoder}
\end{figure}

\begin{figure}[h]
	\includegraphics[width=\columnwidth]{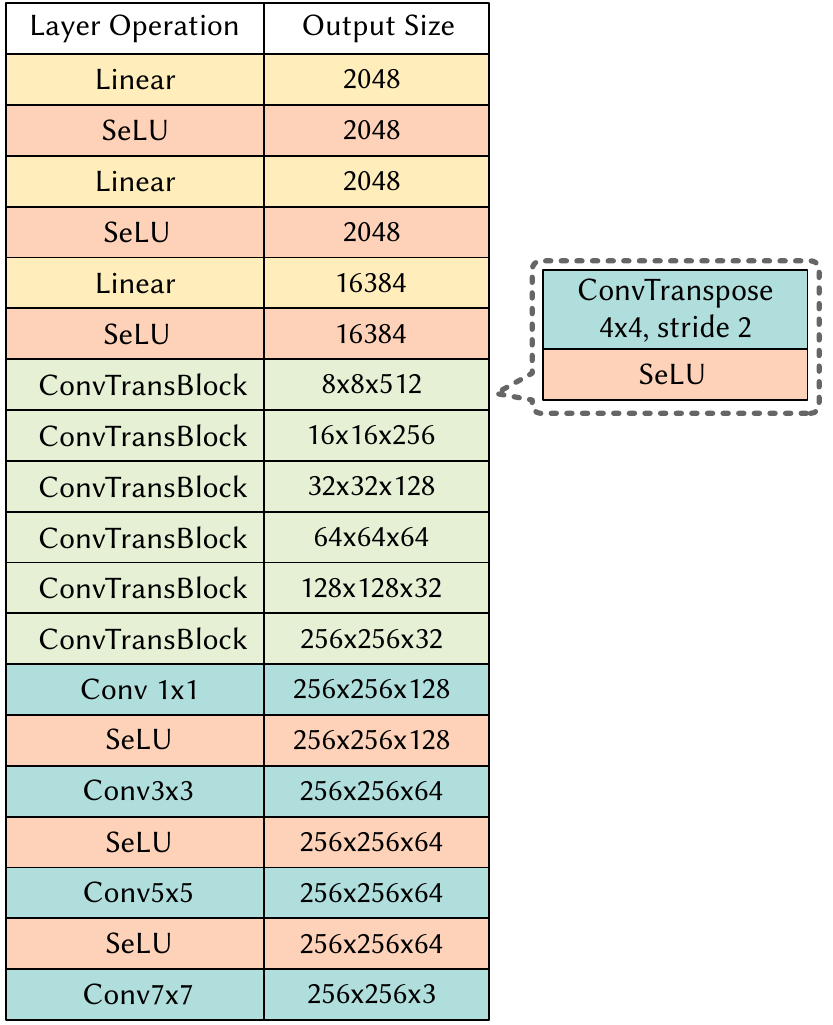}	
	\caption{Decoder architecture}
	\label{fig:decoder}
\end{figure}

\subsubsection{Dataset} \label{dataset}
Our dataset consists of publicly available clips downloaded from YouTube.
The clips contain various types of characters and motion, including "talking head" speech, dancing,
and multi-character interactions.
The clips are pre-processed only by cropping.
For the \emph{John Oliver} clip we compute the minimal square bounding box which contains the face in all frames.
This box is expanded by 15 pixels and used to crop all frames in the sequence.
For all other clips we simply crop the edges to make the frames square.
The cropped frames are then bilinearly downsampled to either $128 \times 128$ or $256 \times 256$px.

\begin{table}
	\centering
	\ra{1.2}
	\setlength{\tabcolsep}{5.2pt}
	\rowcolors{2}{lightbluishgrey}{white}
	\begin{tabular}{lrrrr}
		\toprule
		\rowcolor{white}
		\textit{Clip name} & $\#\mathbf{F}$  & $|\mathbf{Z}|$ & \textit{Resolution} & \textit{Hardware} \\
		John Oliver & 5000 & 200 & $256 \times 256$ & TITAN RTX \\
		Zebra & 5000 & 400 & $128 \times 128$ & GTX 1080Ti \\
		Cookie Monster & 1769 & 250 & $256 \times 256$ & TITAN V \\
		Michelle Obama & 5000 & 200 & $256 \times 256$ & TITAN RTX \\
	\bottomrule
	\end{tabular}
	\smallskip
	\caption{
		Summary of the datasets.
	}\label{tab:data}
\end{table}

\subsection{Interpolation}
\begin{figure*}[htp]
	\includegraphics[width=\textwidth]{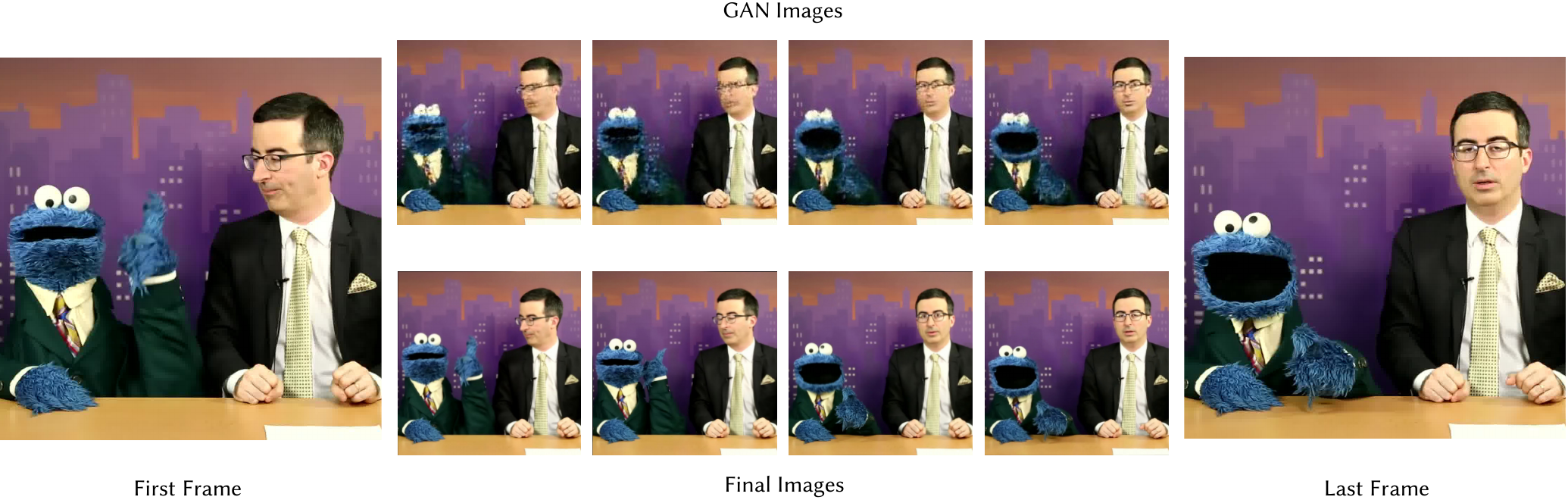}	
	\caption{Interpolating between two random video frames using CHAD. We show the result of our image generator (top) and denoising algorithm (bottom).}
	\label{fig:interpFig}
\end{figure*}

Here we show some results of using CHAD to perform interpolation between two \emph{randomly} chosen key frames from the source video~(\autoref{fig:interpFig}). CHAD is trained separately for each example using one of the datasets from \autoref{tab:data} for each example.

\subsection{Faces}
\begin{figure*}[htp]
	\includegraphics[width=\textwidth]{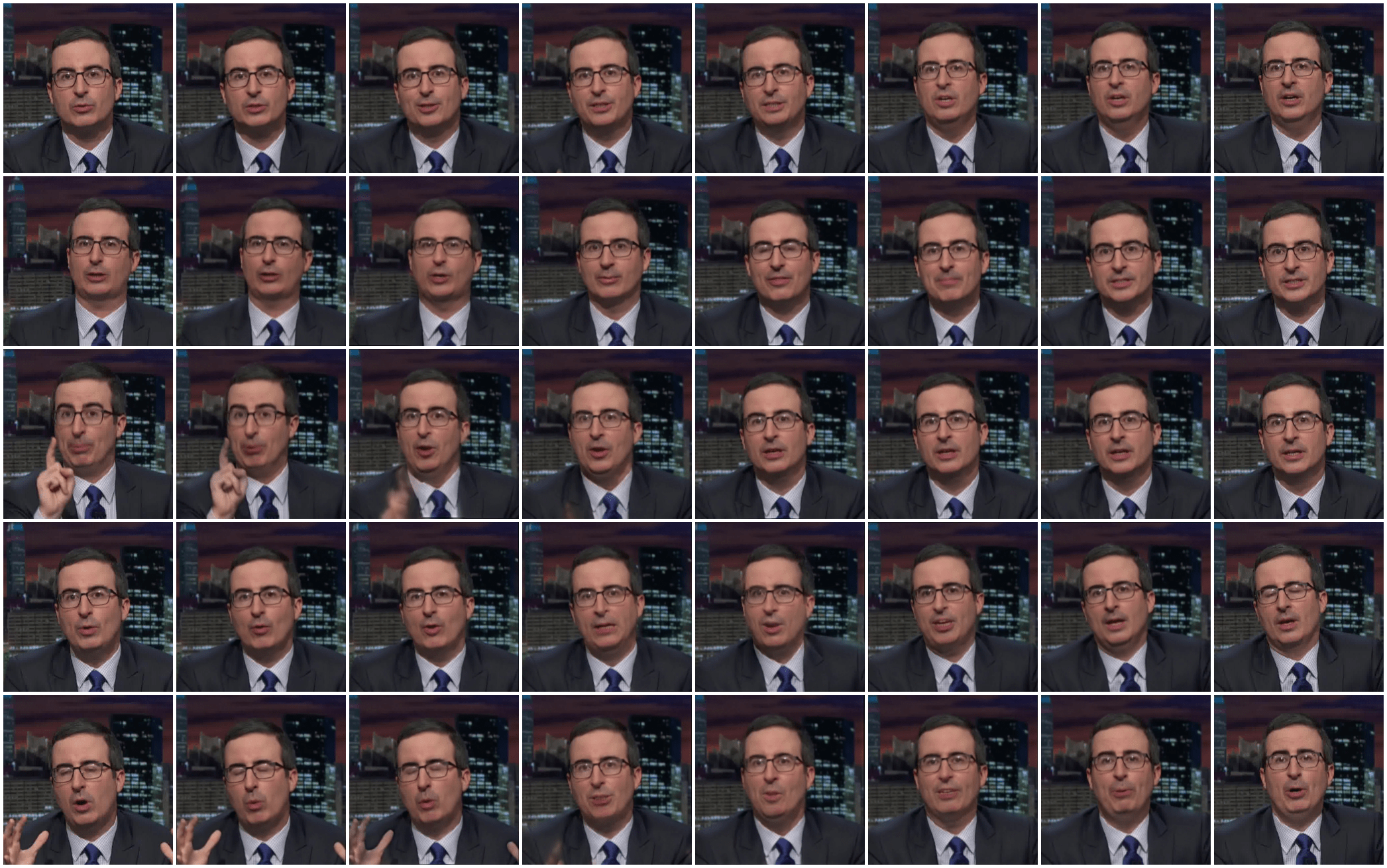}	
	\caption{Examples of interpolating between face poses using CHAD. Left: Initial keyframe. Right: Final keyframe. Middle: Frames synthesized by our method.}
	\label{fig:faceInterp}
\end{figure*}

We begin with some examples of interpolating between facial poses (\autoref{fig:faceInterp}). Because CHAD uses no motion priors and is therefore, not tuned for facial animation, we don't expect it to match the quality of more specialized methods~\cite{kim2018deep}. CHAD's advantage is the ability to synthesize natural frames from only two keyframes. In these examples you will see that CHAD can adequately generate suitable facial motion like blinking and can interpolate hand motion (even when the hand enters and exits the frame). To the authors knowledge, CHAD is the first algorithm to be able to perform synthesis of this detail by purely relying on input data. 

\subsection{Whole Body}
\begin{figure*}[htp]
	\includegraphics[width=\textwidth]{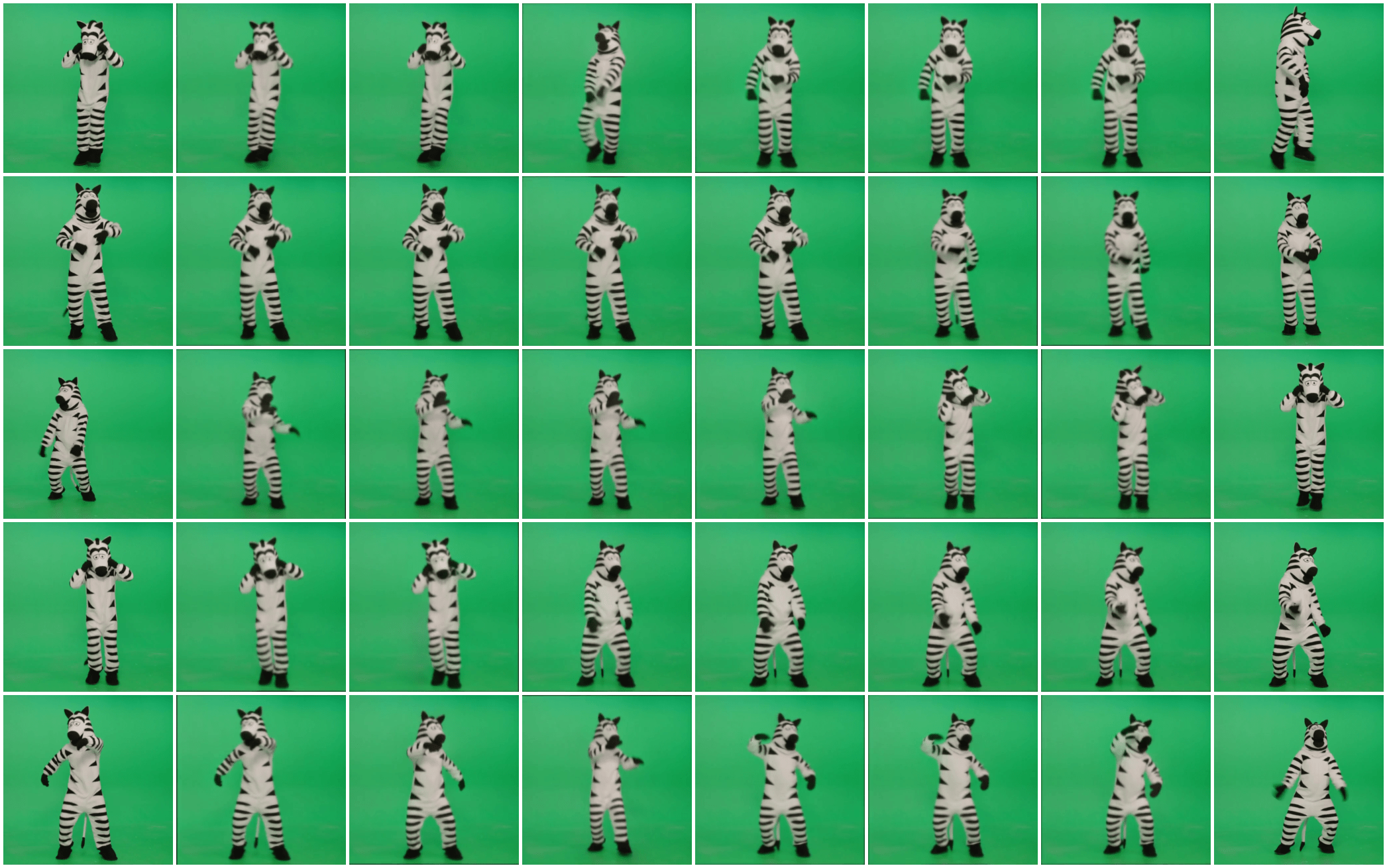}	
	\caption{Examples of interpolating whole body poses using CHAD. Left: Initial keyframe. Right: Final keyframe. Middle: Frames synthesized by our method.}
	\label{fig:wholeBody}
\end{figure*}

CHAD can also interpolate between whole body keyframes without any adjustments to the network architecture~(\autoref{fig:wholeBody}). Here we show the results of performing animation synthesis on a bipedal zebra dataset taken from youtube. Notice that CHAD is capable of interpolating between poses with different facing and large limb motion. 

\subsection{Multiple Characters} 
\begin{figure*}[htp]
	\includegraphics[width=\textwidth]{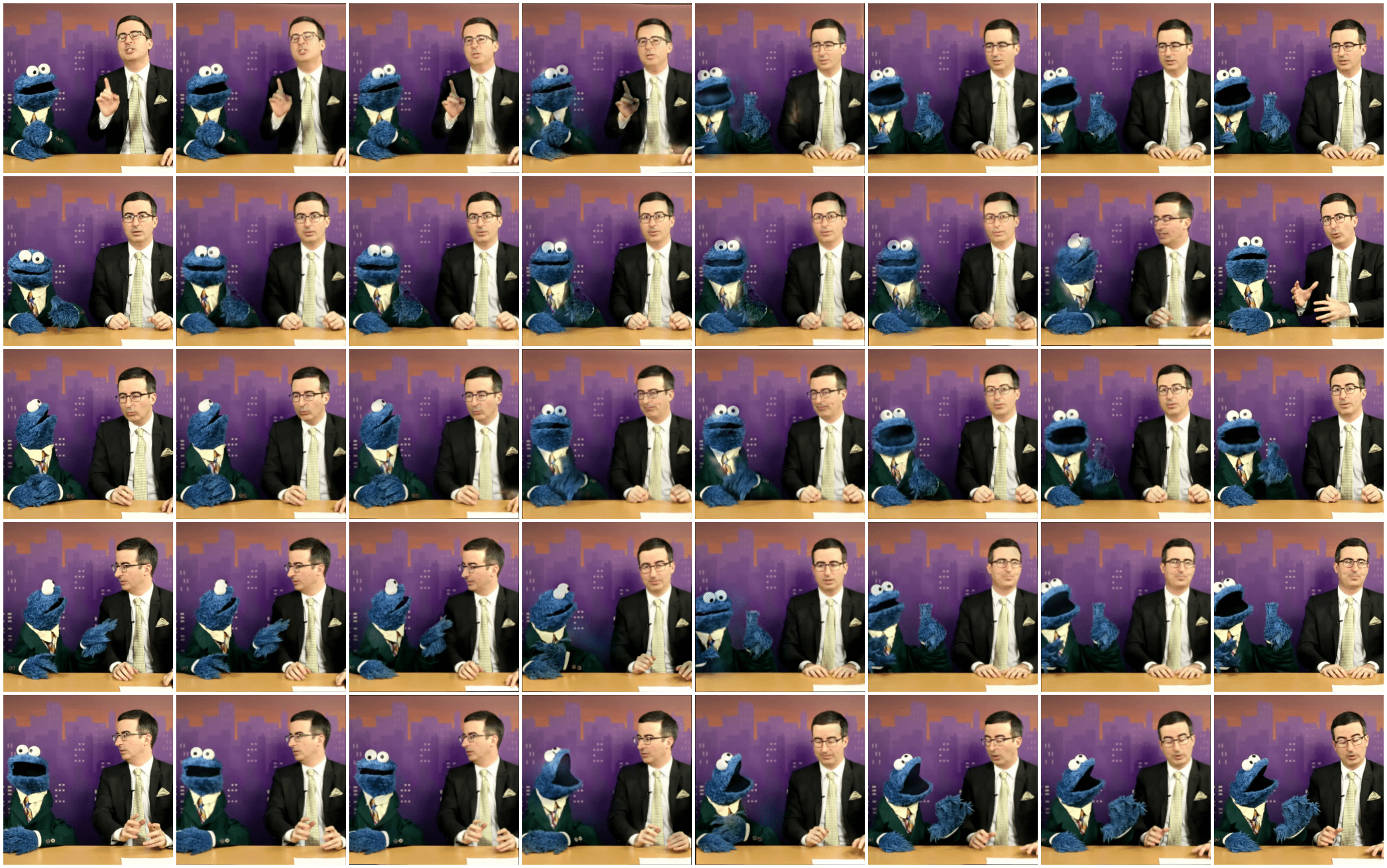}	
	\caption{Examples of interpolating a two character poses using CHAD. Left: Initial keyframe. Right: Final keyframe. Middle: Frames synthesized by our method.}
	\label{fig:multiple}
\end{figure*}

Finally, in order to stress the flexibility of CHAD we demonstrate interpolation between random frames of a video containing two characters~(\autoref{fig:multiple}). Again we CHAD is able to synthesize natural motion for both characters that is consistent with the input video. Of particular interest is the ``googliness'' of the blue monster's eyes. 


%% file: conclusions.tex
\section{Conclusions and Future Work}
\label{sec:conclusion}
In this paper we have presented CHAD, our machine learning approach to creating image-based animation of humanoid characters driven  by sparse temporal input. CHAD's strength is its generality. By eschewing any prior models about character animation, CHAD can produce a wider range of animated motions than previous approaches. This flexibility is enabled by our novel configuration manifold learning approach and our new asymmetric architecture which we have justified both theoretically and experimentally. 

CHAD significantly lowers the barrier of entry for creating image-based animations, requiring only a single short to medium length youtube video to learn a model of humanoid motion. We also believe that CHAD opens up a number of avenues for exciting future work. 

First, our current image denoising approach is less than perfect. It generates acceptably sharp images at the cost of a reduction in temporal smoothness. It can also introduce some ghosting when the source and target frames for detail transfer don't align perfectly. This is a shame because our GAN images, despite lacking detail occasionally, capture interpolated motion extremely well. Ideally, we would be able to retain the temporal coherence evidenced in our output GAN frames, however this seems out of reach without resorting to extremely long runtimes and large amounts of data~\cite{karras2018progressive}. Using our image synthesis procedure to perform on-the-fly data augmentation during GAN training may help us overcome some of these challenges.

Second, we would also like to extend CHAD from the 2D domain to the 3D domain by attempting to reconstruct  configuration manifolds for 3D objects using depth scan data from commodity hardware such as the iPhone X. Such an algorithm could lower the barrier of entry for 3D animation in the same way we feel that CHAD has done for 2D animation. 

Finally, we are curious about using our learned configuration spaces for autonomous character animation. Keyframes could be used as the state for an animation controller with actions being transitions between frames. CHAD provides a data-driven means to generate poses between these discrete states and could serve to bring image-based autonomous actors to live. In order to enable this and other explorations we intend to release the CHAD source code and pre-trained networks for all examples shown in this submission.